\documentclass[twocolumn,showpacs,unsortedaddress,superscriptaddress,showkeys,preprintnumbers,letterpaper,nofootinbib]{revtex4-1}
\usepackage{soul}
\usepackage{acronym}
\usepackage{amsmath}
\usepackage{amssymb}
\usepackage{mathrsfs}
\usepackage{graphicx}
\usepackage{float}
\usepackage{appendix}
\usepackage{comment}
\usepackage{subfigure}
\usepackage[usenames]{color}
\usepackage[normalem]{ulem}
\usepackage[bookmarksnumbered, bookmarksopen, breaklinks, colorlinks]{hyperref}
\pdfoutput=1

\newcommand{\AEI}{Albert-Einstein-Institut, Max-Planck-Institut f\"ur
Gravitationsphysik, D-30167 Hannover, Germany}
\newcommand{\Leibniz}{Leibniz Universit\"at Hannover, D-30167 Hannover, Germany}

\newcommand{\cross}{\times}
\newcommand{\RA}{\ensuremath{\alpha}}
\newcommand{\dec}{\ensuremath{\delta}}
\newcommand{\A}{\ensuremath{\mathcal{A}}}
\newcommand{\D}{\ensuremath{\mathcal{D}}}
\newcommand{\F}{\ensuremath{\mathcal{F}}}
\newcommand{\G}{\ensuremath{\mathcal{G}}}
\newcommand{\M}{\ensuremath{\mathcal{M}}}

\newcommand{\Y}{\ensuremath{Y}}
\newcommand{\mat}[1]{{\boldsymbol #1}}

\newcommand*{\diff}{\,\mathrm{d}}

\newcommand{\abs}[1]{\left\lvert #1 \right\rvert}

\begin{document}


\title{Metrics for multi-detector template placement in searches for
short-duration nonprecessing inspiral gravitational-wave signals}

\author{Drew~Keppel}  
\email{drew.keppel@ligo.org}
\affiliation{\AEI}
\affiliation{\Leibniz}

\begin{abstract}
Using the family of multi-detector \F-statistic metrics for short duration,
nonprecessing inspiral signals, we derive a marginalized metric that is
directly applicable to the problem of generating template banks for coincident
and coherent multi-detector searches for gravitational-waves. This metric is
compared to other average metrics, such as that proposed for the case of
searches associated with continuous signals from rotating neutron stars. We
show how the four-dimensional metric can be separated into two two-dimensional
metrics associated with the sky and mass parameter subspaces, allowing the
creation of separate template banks for these subspaces. Finally, we present an
algorithm for computing the mass space metric associated with both coincident
and coherent multi-detector targeted or all-sky searches for short duration,
nonprecessing inspiral gravitational-wave signals.
\end{abstract}

\maketitle

\acrodef{BNS}{binary neutron star}
\acrodef{CBC}{compact binary coalescence}
\acrodef{GW}{gravitational-wave}
\acrodef{GRB}{gamma-ray burst}
\acrodef{PN}{post-Newtonian}
\acrodef{PSD}{power spectral density}
\acrodef{ROC}{Receiver Operator Characteristic}
\acrodef{SNR}{signal-to-noise ratio}
\acrodef{SPA}{stationary phase approximation}
\acrodef{SVD}{singular value decomposition}
\acrodef{IMR}{inspiral-merger-ringdown}
\acrodef{ASD}{amplitude spectral density}

\section{Introduction}

Advanced versions of the broadband \ac{GW} detectors, such as
LIGO~\cite{Harry:2010zz}, GEO~\cite{Willke:2006uw, GEO2010},
KAGRA~\cite{Somiya:2011np}, and Virgo~\cite{advVirgo2009}, are currently being
built and installed. Once the detectors start operating and reach their design
sensitivity, they are expected to be sensitive to detect \ac{GW} signals from
tens of \ac{BNS} systems per year~\cite{rates2010} with $\sim$10\% localized to
better than 5 $\textrm{deg}^2$~\cite{localization2013}.

In order to search \ac{GW} data from a network of detectors for these signals,
banks of filters, called ``template banks", are used to matched-filter the
data. Each filter is constructed using the signal waveform associated with a
specific choice of signal parameters. There exists a variety of methods that
choose which points in parameter space should be used to generate the filters
of a template bank~\cite{Owen:1995tm, Cokelaer2007, Prix2007, Harry2009,
Messenger2009, Manca2010}, however the most common method involves computing a
metric on the parameter space.  This metric describes the fractional loss of
signal strength as a function of the mismatch between the parameters of a
template waveform and the parameters of an observed
signal~\cite{Sathyaprakash:1991mt, Owen:1995tm}.  Previous works have reported
the calculation of metrics that can be used for template bank construction for
single detector analyses~\cite{Owen:1995tm, Owen:1998dk, Brown:2012qf,
Keppel:2013kia}, however only families of metrics have been reported for
network analyses~\cite{Pai:2000zt, Keppel:2012ye}.

Combining the results from a network of detectors can be done in multiple ways.
One approach is to analyze data from different detectors separately, generating
single-detector ``triggers" associated with peak values and times of a single
detector's \ac{SNR} time series, before looking for coincident triggers between
multiple detectors~\cite{Brown:2005zs, Babak:2012zx}. If different template
banks are used to analyze data from different detectors, triggers from
different detectors associated with different points in parameter space will
need to be somehow combined. This combination can increase the false-alarm
probability associated with a fixed sensitivity, which motivates using the same
template bank to analyze all detectors in a particular network. Another
approach is to combine data from multiple detectors in a coherent
manner~\cite{Bose:1999bp, Bose2000, Pai:2000zt, Finn2001, Cutler2005,
Harry2011}, producing triggers associated with the detector network.

In this work, we present an algorithm for how the $\F$-statistic metric
family~\cite{Keppel:2012ye} can be marginalized over physical parameters in
order to obtain a metric that can be used for template bank construction for
multiple-detector analyses. This paper is organized as follows. In
Section~\ref{sec:params}, we present the parameters that describe
short-duration, nonprecessing inspiral \ac{GW} signals and how they can be
grouped.  In Section~\ref{sec:prelim}, we recall the construction of the
$\F$-statistic and its metric. In Section~\ref{sec:fstatavemetric}, we discuss
possible averaging procedures one could use in order to obtain a metric that is
independent of the amplitude parameters. This is necessary in order to
construct template banks covering the parameter space that must be searched in
a templated manner.  In Section~\ref{sec:metricsep}, we show how the
four-dimensional metric on the mass and sky parameter space can be separated
into a two-dimensional metric for the mass subspace and a two-dimensional
metric for the sky subspace. Section~\ref{sec:massmetric} describes the
construction of a sky-parameter independent metric for the mass subspace, which
is shown to be valid for both coherent and coincident network analyses.
Section~\ref{sec:skymetric} describes the construction of a metric for the sky
subspace. The major differences between a previously derived average
metric~\cite{PrixFstatMetric} and the marginalized metric presented here are
shown to be associated with the network response power to the subdominant
polarization.

\section{Parameters}
\label{sec:params}

The parameters characterizing non-spinning short-duration inspiral signals can
be grouped into several disjoint sets. Intrinsic parameters are those
parameters that affect the phase evolution and overall structure of the
waveform. For binaries where the component objects' spins can be neglected,
these are the mass parameters $\{\eta, \M_c\}$, where $\eta := m_1 m_2 / (m_1 +
m_2)^2$ is the symmetric mass ratio, $\M_c := (m_1 + m_2) \eta^{3/5}$ is the
chirp mass, and $m_1$ and $m_2$ are the component masses of the objects in the
binary. Extrinsic parameters are those which affect the amplitude, time of
arrival, or phase of the signal seen by \ac{GW} detectors. These are given by
$\{\D, \phi_0, \psi, \iota, \RA, \dec, t_c\}$, where $\D$ is the distance to
the source (the extrinsic amplitude is related to the distance by $h_0 \propto
1/\D$), $\phi_0$ is the phase offset for the waveform, $\psi$ is the
polarization angle between the source and detector frames, $\iota$ is the
inclination angle between the line of sight and the orbital angular momentum
vector, $\RA$ is the right ascension, $\dec$ is the declination, and $t_c$ is
the time of coalescence at the geocenter.

The extrinsic parameters $\{\D, \phi_0, \psi, \iota\}$ are seen to only affect
the amplitude and phase offset of the signal based on how they enter the signal
model, which will be explained more completely in Sect.~\ref{sec:prelim}. As
discussed in~\cite{Keppel:2012ye}, and references therein, the maximum
likelihood estimate of these parameters can be measured analytically through
the construction of the \F-statistic. In addition, the coalescence time $t_c$
can be found efficiently through the use of the Fast Fourier Transform used to
perform the matched-filtering in constructing the individual detector \ac{SNR}
time series.

The remaining parameters, two extrinsic parameters associated with the sky
coordinates $\{\RA, \dec\}$ and two intrinsic mass parameters, must be searched
with an explicit loop over different templates.

\section{Preliminaries}
\label{sec:prelim}

Recalling from \cite{Keppel:2012ye}, we model a \ac{GW} signal from
short-duration nonprecessing inspiralling compact objects seen by detector $\Y$
using linear combinations of four detector-dependent
\emph{polarization-weighted basis waveforms},
\begin{equation}
s^Y = \sum_{\mu} \A^\mu h_{\mu}^{\Y},
\end{equation}
where $\{\A^\mu\}$ are the amplitude parameters, which are given in
Appendix~\ref{app:ampparam}, and $\{h_{\mu}^{\Y}\}$ are explicitly given in
Appendix~\ref{app:basiswaveforms}. From this model of the signal, the maximum
likelihood ratio estimate of the signal can be found using the
\F-statistic~\cite{Cutler2005},
\begin{equation}\label{eq:fstat}
\F := \ln \Lambda(\mat{x}; \mat{s}_{\rm ML}) = \frac{1}{2} x_\mu \M^{\mu \nu}
x_\nu.
\end{equation}
In this expression, the quantity $x_\mu := (\mat{x}|\mat{h}_\mu)$ is a vector
inner product of detector data $x^{\Y}$ with waveform $h_\mu^{\Y}$. The matrix
component $\M^{\mu \nu}$ is the $\mu,\nu$ component of the inverse of $\M_{\mu
\nu}$, where $\M_{\mu \nu} := (\mat{h}_\mu|\mat{h}_\nu)$. In these quantities,
the vector inner product is defined to be $(\mat{a}|\mat{b}) := \sum_{\Y}
(a^\Y|b^\Y)$ and the inner product between two waveforms associated with
detector $\Y$ is defined as
\begin{equation}\label{eq:inner}
(x^\Y|y^\Y) := 4 \Re \int \frac{\tilde{x}^\Y(f) \tilde{y}^{\Y*}(f)}{S^{\Y}(f)}
df,
\end{equation}
with the operator $\Re$ extracting the real part of its argument and
$S^{\Y}(f)$ representing the one-sided \ac{PSD} of the noise in detector $\Y$.
To be explicit, $\M_{\mu\nu}$ contains the plus-polarization network response
power, $A := (\mat{h}_1|\mat{h}_1)$, the cross-polarization network response
power, $B := (\mat{h}_2|\mat{h}_2)$, and the mixed-polarization network
response power, $C := (\mat{h}_1|\mat{h}_2)$. It takes the form
\begin{equation}\label{eq:M}
\M^{\mu\nu} = \left(\begin{array}{cccc}
A & C & 0 & 0 \\
C & B & 0 & 0 \\
0 & 0 & A & C \\
0 & 0 & C & B \end{array}\right)_{\mu \nu}.
\end{equation}
The inverse matrix $\M^{\mu\nu}$ then takes the form
\begin{equation}
\M^{\mu\nu} = \frac{1}{D}\left(\begin{array}{cccc}
B & -C & 0 & 0 \\
-C & A & 0 & 0 \\
0 & 0 & B & -C \\
0 & 0 & -C & A \end{array}\right)^{\mu \nu}
\end{equation}
with
\begin{equation}\label{eq:D}
D := AB - C^2.
\end{equation}

It has previously been shown that the normalized projected Fisher matrix can be
used as a metric for the \F-statistic associated with short-duration
nonprecessing inspiral signals. For short-duration nonprecessing inspiral
signals, this metric is of the form
\begin{equation}\label{eq:fstatmetric}
g^{\F}_{i j} = \frac{\A^\alpha \G_{\alpha \beta i j} \A^\beta}{\A^\alpha
\M_{\alpha \beta} \A^\beta}.
\end{equation}
where the projected Fisher matrix $\G_{\mu \nu i j}$ has the following
components,
\begin{equation}\label{eq:mismatches}
\G_{\mu \nu i j} = \left(\begin{array}{cccc}
m^1_{ij} & m^3_{ij} & 0 & m^4_{ij} \\
m^3_{ij} & m^2_{ij} & -m^4_{ij} & 0 \\
0 & -m^4_{ij} & m^1_{ij} & m^3_{ij} \\
m^4_{ij} & 0 & m^3_{ij} & m^2_{ij} \end{array}\right)_{\mu \nu},
\end{equation}
with the Greek indices referring to the enumerated amplitude parameters and the
Latin indices referring to the extrinsic parameters $\{\RA, \dec, t_c\}$ as
well as the intrinsic parameters.  The formulae for these mismatch components
$m^k_{ij}$ can be found in Section V of \cite{Keppel:2012ye}. It is useful to
note that the mismatch components are independent of the amplitude parameters.

From \eqref{eq:fstatmetric}, it is apparent that although the amplitude
parameters have been projected out of the Fisher matrix, the \F-statistic
metric is still dependent on the amplitude parameters. It is for this reason
that \eqref{eq:fstatmetric} is said to represent a family of metrics.

\section{Average Metrics}
\label{sec:fstatavemetric}

In this section, we outline two different approaches toward obtaining an
``average" metric that is independent of the amplitude parameters. The first
method is based on finding the extrema of the possible mismatches, while the
second is based on marginalization over the physical parameters that comprise
the amplitude parameters.

\subsection{Extrema-based average metric}
\label{subsec:exavemetric}

The average metric that has previously been derived for the \F-statistic is
based on the midpoint of the range of possible mismatches, which we shall call
the \emph{extrema-averaged metric}.  In Ref.~\cite{PrixFstatMetric} it was
shown how the extrema of possible mismatch for arbitrary values of the
amplitude parameters can be obtained. Explicitly, for a metric of the form in
\eqref{eq:fstatmetric}, the extrema $\widehat{m}_\F$ can be found as the
eigenvalues of $\M^{\mu \alpha} \G_{\alpha\nu} \Delta\lambda^i
\Delta\lambda^j$.  The two independent eigenvalues of this matrix are given by
\begin{equation}
\widehat{m}_{\F}^{\mathrm{max}|\mathrm{min}} = \overline{m}_\F \pm
\sqrt{\overline{m}_\F^2 - \widetilde{m}^2},
\end{equation}
where the average value $\overline{m}_\F$ is given by
\begin{equation}\label{eq:avemismatchext}
\overline{m}_\F = (2D)^{-1} (B m^1 + A m^2 - 2C m^3),
\end{equation}
and the spread $\widetilde{m}^2$ is given by
\begin{equation}
\widetilde{m}^2 = D^{-1} ( m^1 m^2 - m^3 m^3 - m^4 m^4).
\end{equation}
For simplicity in the above equations, we have used $m^k := m^k_{ij}
\Delta\lambda^i \Delta\lambda^j$.  Based on the definition of the average
mismatch in \eqref{eq:avemismatchext}, the average metric can be constructed
from the mismatch components as
\begin{equation}\label{eq:avefstatmetric}
\overline{g}^{\F}_{i j} = (2D)^{-1}(B m^1_{ij} + A m^2_{ij} - 2C m^3_{ij}).
\end{equation}

This formulation of the average metric is also useful when one wants a metric
that has been maximized over the amplitude parameters, as is suggested
in~\cite{Pai:2000zt}. Since $\overline{g}^{\F}_{i j}$ is the average of the
minimum and maximum possible mismatches, the maximum possible mismatch will be
bounded by $2\overline{g}^{\F}_{i j}$~\cite{PrixFstatMetric}.

\subsection{Marginalization-based average metric}
\label{subsec:margavemetric}

An alternative approach to computing an average metric is to marginalize the
metric over the extrinsic parameters that comprise the amplitude parameters
(i.e., the polarization angle, inclination angle, phase angle, and distance).
To do this, we will need to know the probability distribution functions
associated with those parameters.

The polarization and phase angles can vary between $0$ and $2\pi$ and
physically have no preferred value. Thus uniform distributions in those
variables are the correct distributions.  Recalling that the inclination angle
is the angle between the orbital angular momentum vector of the source and the
line of sight, which can take on values between $0$ and $\pi$, we find that
this variable is distributed uniformly on a sphere. Thus, the correct
distribution for this variable is uniform in the cosine of the angle. Finally,
as sources are assumed to be uniformly distributed in space\footnote{This of
course neglects stellar evolution and cosmological effects, the latter of which
would be non-negligible for the larger binary black hole sources to which the
advanced detectors will be sensitive.}, we expect to see relatively more
signals further away with a distribution proportional to the square of the
distance. To summarize, the physical distributions for variables that enter the
amplitude parameters are
\begin{align}
P(\psi) &\propto 1, \\ P(\phi_0) &\propto 1, \\ P(\iota) &\propto \sin \iota,
\\ P(\D) &\propto \D^2.
\end{align}

Let us start the averaging procedure by examining how the distance enters into
the metric in \eqref{eq:fstatmetric}. From \eqref{eq:M} and
\eqref{eq:mismatches}, we see that $\M$ and $\G_{ij}$ are independent of the
amplitude parameters. Since the extrinsic amplitude, and thus the distance,
enters the polarization amplitudes of \eqref{eq:polamps} in a uniform manner,
it can be pulled out of the amplitude parameters, \eqref{eq:ampparams}, that
appear in the metric. Due to equal powers of amplitude parameters in the
numerator and denominator of the metric, we come to the conclusion that the
metric is actually independent of the distance. Marginalizing the metric over
the distance then becomes the simple problem of integrating $P(\D)$ over $\D$
where we need to decide the proper choice for the upper bound
$\D_\mathrm{max}$,
\begin{equation}\label{eq:metricvoldistmarg}
\langle g^\F_{ij} \rangle_\D \propto \int_0^{\D_\mathrm{max}} g^\F_{ij} P(\D) \diff\D =
\frac{g^\F_{ij}}{3} \D_\mathrm{max}^3.
\end{equation}
The result of this is a weight that will be useful in further marginalization
over other variables.

At a fixed value of the coherent \ac{SNR} $\rho_*$, we will be able to see out
to a distance given by $\D_* = \sqrt{(s|s)} / \rho_* = \sqrt{\A^\alpha
\M_{\alpha \beta} \A^\beta}/\rho_*$. This represents the density of sources a
network of detectors would be sensitive to for a given solid angle on the sky,
$\mathrm{d}\Omega$, with $\rho \geq \rho_*$. This is the distance we will use
as the upper cutoff for the distance marginalization and it is evidently
dependent on $\iota$ and $\psi$.

Marginalizing over the other extrinsic parameters that make up the amplitude
parameters leaves us with
\begin{align}\label{eq:metricvolfullmarg}
\langle g^\F_{ij} \rangle_{\D,\psi,\phi_0,\iota} \propto& \int_0^{2\pi}
\int_0^{2\pi} \int_0^{\pi} \frac{g^\F_{ij}}{3} \left(\frac{\sqrt{\A^\alpha
\M_{\alpha \beta} \A^\beta}}{\rho_*}\right)^3 \nonumber \\
& \quad  P(\phi_0) P(\psi) P(\iota) \diff\psi \diff\phi_0 \diff\iota \nonumber
\\
\propto& \frac{1}{3\rho_*^3} \int_0^{2\pi} \int_0^{2\pi} \int_0^{\pi} \sin\iota
\nonumber \\
& \quad \A^\alpha \G_{\alpha \beta i j} \A^\beta \sqrt{\A^\delta \M_{\delta
\gamma} \A^\gamma} \diff\psi \diff\phi_0 \diff\iota.
\end{align}
In order to fix the multiplicative factor so that the probability distributions
are normalized, we compute
\begin{align}\label{eq:normvolfullmarg}
N :=& \int_0^{\D_\mathrm{max}} \int_0^{2\pi} \int_0^{2\pi} \int_0^{\pi} P(\D)
P(\phi_0) P(\psi) P(\iota) \nonumber \\
&\quad \diff\D \diff\psi \diff\phi_0 \diff\iota \nonumber\\
=& \frac{1}{3 \rho_*^3} \int_0^{2\pi} \int_0^{2\pi} \int_0^{\pi} \sin\iota
\left(\A^\alpha \M_{\alpha \beta} \A^\beta\right)^{3/2} \diff\psi \diff\phi_0
\diff\iota.
\end{align}
Unfortunately this is are far as we can take the analytic treatment of the
marginalization process. Instead, let us consider a more convenient, though
less optimal, choice for the distance distribution. If we assume $P(\D) \propto
\D$, \eqref{eq:metricvoldistmarg} then becomes
\begin{align}\label{eq:metricdistmarg}
\langle g^\F_{ij} \rangle_\D &\propto \int_0^{\D_\mathrm{max}} g^\F_{ij} P(\D)
\diff\D \nonumber\\
&\propto \frac{1}{2\rho_*^2} \A^\alpha \G_{\alpha \beta i j} \A^\beta.
\end{align}
Continuing with marginalizing the other variables, and remembering $\G_{\mu \nu
i j}$ is independent of the amplitude parameters, \eqref{eq:metricvolfullmarg}
becomes
\begin{multline}\label{eq:metricfullmarg}
\langle g^\F_{ij} \rangle_{\D,\psi,\phi_0,\iota} \propto  \\
\frac{1}{2\rho_*^2} \G_{\alpha \beta i j} \int_0^{2\pi} \int_0^{2\pi}
\int_0^{\pi} \A^\alpha \A^\beta \sin\iota \diff\psi \diff\phi_0 \diff\iota.
\end{multline}
In a similar manner, the normalization constant becomes
\begin{align}\label{eq:normfullmarg}
N = \frac{1}{2 \rho_*^2} \M_{\alpha \beta} \int_0^{2\pi} \int_0^{2\pi}
\int_0^{\pi} \A^\alpha \A^\beta \sin\iota \diff\psi \diff\phi_0 \diff\iota.
\end{align}
To complete the computation of both \eqref{eq:metricfullmarg} and
\eqref{eq:normfullmarg}, all that is needed are the integrals over products of
amplitude parameters, which we compute explicitly in
Appendix~\ref{app:ampaves}.  The only non-zero integrated amplitude parameter
combinations are associated with the diagonal elements of $\M_{\mu\nu}$ and
$\G_{\mu \nu i j}$. In addition, the coefficients coming from those integrals
\eqref{eq:ampfullint} are identical. This leads to the result that the
approximate marginalized metric of \eqref{eq:metricfullmarg} is 
\begin{equation}
\langle g^\F_{ij} \rangle_{\D,\psi,\phi_0,\iota} = \frac{m^1_{ij} +
m^2_{ij}}{A+B}.
\end{equation}
In the remainder of this document, unless otherwise noted, this is what we mean
when we refer to the marginalized metric.

It should be noted that this expression for the marginalized metric holds for
both the situation where signals are assumed to come from a randomly oriented
distribution of binaries (e.g., all-sky searches) and the situation where the
signals are assumed to come from binaries whose orbital plane is perpendicular
to the line of sight (e.g., searches targeting signals from short \ac{GRB}
progenitors).

\section{Separation of Mass and Sky Dimensions}
\label{sec:metricsep}

In this section we look at the correlations that are present in the metric
between the mass and sky parameters. In general, these correlations are small,
an example of which can be seen in Tables~\ref{tab:corrs} and
\ref{tab:avecorrs}. Therefore, the four-dimensional mass-sky metric can be
approximately separated into two two-dimensional metrics covering the sky
parameter space and the mass parameter space separately. This implies that
template banks can be constructed separately for each of these subspaces. This
is a desirable property because it allows the computation of the coherent
\ac{SNR} to take place in two stages~\cite{Pai:2000zt}. First, the
single-detector data can be match-filtered to produce single-detector \ac{SNR}
time series for each mass template, an operation that involves convolving the
template waveform with the data, weighted by the detector's inverse \ac{PSD}.
These time series can then be appropriately shifted, weighted, and combined in
order to form the coherent \ac{SNR} for many different sky positions. See
\cite{Pai:2000zt} and \cite{DalCanton2013} for a more details accounting of the
computational costs of performing coherent searches.

\begin{table}
\begin{tabular}{c|c|c|c|c|}
 & \RA & \dec & $\M_c$ & $\eta$ \\ \hline
\RA & $1$ & $-0.868$ & $-0.0104$ & $-0.0826$ \\ \hline
\dec & $-0.868$ & $1$ & $0.0148$ & $0.0775$ \\ \hline
$\M_c$ & $-0.0104$ & $0.0148$ & $1$ & $-0.873$ \\ \hline
$\eta$ & $-0.0826$ & $0.0775$ & $-0.873$ & $1$ \\ \hline
\end{tabular}
\caption{Here we show the correlations between the sky and mass dimensions of
the metric associated with the HLV network for randomly chosen values of the
sky-position and physical parameters that enter the amplitude parameters:
$\RA=2.98$, $\dec=0.617$, $\cos\iota = -0.166$, $\psi=-1.58$, $\phi_0=-0.967$.}
\label{tab:corrs}\end{table}
%

\begin{table}
\begin{tabular}{c|c|c|c|c|}
 & \RA & \dec & $\M_c$ & $\eta$ \\ \hline
\RA & $1$ & $0.443$ & $0.0203$ & $0.0657$ \\ \hline
\dec & $0.443$ & $1$ & $0.0172$ & $0.0597$ \\ \hline
$\M_c$ & $0.0203$ & $0.0172$ & $1$ & $0.864$ \\ \hline
$\eta$ & $0.0657$ & $0.0597$ & $0.864$ & $1$ \\ \hline
\end{tabular}
\caption{Here we show the average absolute value correlations between the sky
and mass dimensions of the metric associated with the HLV network. The average
is computed by marginalizing these correlations over the physical parameters
that enter the amplitude parameters and the sky-location parameters. The
cross-correlations between the mass and sky parameter subspaces are about an
order of magnitude smaller than the cross-correlations within those subspaces.}
\label{tab:avecorrs}\end{table}

\subsection{Mass metric}
\label{sec:massmetric}

Once the marginalized metric is split between the mass and sky parameter
subspaces, we see that this mass metric is still dependent on the sky-location
parameters. One way to address this would be to marginalize this metric over
the physical distributions associated with the sky-location parameters.
Unfortunately, this will result in integrals that cannot be computed
analytically.

An alternative approach obtained by looking at the coherent \ac{SNR}
marginalized over the angle parameters that enter the amplitude parameters and
the sky-location parameters. When the parameters of the filter waveform match
those of the signal, the coherent \ac{SNR} is given by
\begin{equation}
\rho^2 = \A^\alpha \M_{\alpha \beta} \A^\beta.
\end{equation}
Marginalizing this over the angle parameters will result in an integral similar
to that of the normalization constant of \eqref{eq:normfullmarg}, leading to
\begin{equation}\label{eq:partialsnrmarg}
\langle \rho^2 \rangle_{\psi, \phi_0, \iota} = \frac{2 h_0^2}{5} \sum_\Y
\left(F_+^\Y F_+^\Y + F_\times^\Y F_\times^\Y\right) (h^\Y|h^\Y),
\end{equation}
where $h^\Y$ can be either the cosine or the sine waveform. Let us look at how
the sky-location parameters enter these quantities. The detector responses are
each functions of the sky-location parameters. In general, the waveforms $h^\Y$
are also dependent on the sky-location parameters through a time offset.
However, in the combination $(h^\Y|h^\Y)$ this dependence disappears.
Marginalizing this combination of detector responses over the sky-location
parameters results in
\begin{equation}
\left\langle F_+^\Y F_+^\Y + F_\times^\Y F_\times^\Y
\right\rangle_{\alpha,\delta} = \frac{2}{5}.
\end{equation}
This implies that after marginalizing over the sky-location parameters,
\eqref{eq:partialsnrmarg} will be given by
\begin{equation}
\langle \rho^2 \rangle_{\phi,\psi,\cos\iota,\alpha,\delta} = \frac{4 h_0^2}{25}
\sum_\Y (h^\Y|h^\Y).
\end{equation}
The sum over detectors of $(h^\Y|h^\Y)$ can be represented by a single inner
product associated with a virtual detector whose \ac{PSD} is chosen to be that
of the harmonic sum of the original detectors' \acp{PSD}.

\begin{figure}
\includegraphics[]{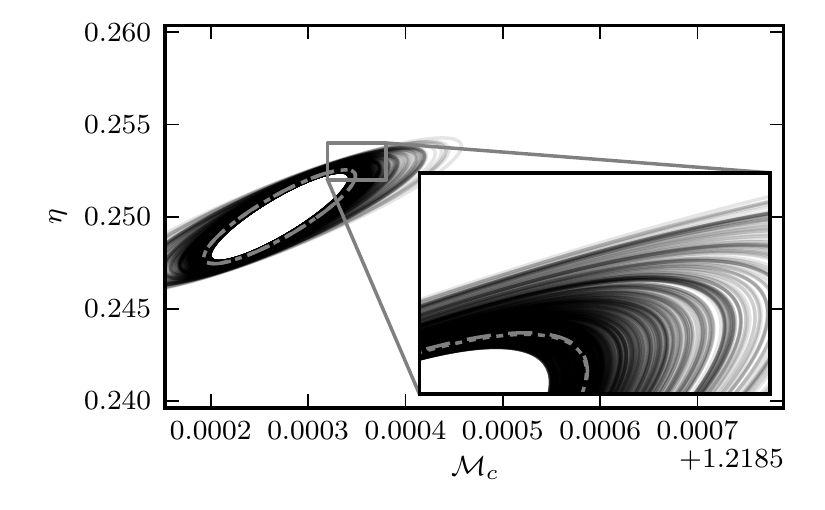}
\caption{We show 3\% mismatch ellipses associated with different mass metrics
for a single point in the mass parameter space. The many transparent ellipses
come from different choices of the sky location, polarization, and inclination
angle, each chosen randomly with the appropriate distribution. The dashed
ellipse shows the metric numerically-marginalized using a $P(\D) = \D^2$
distribution, and projected down to the mass subspace.  The dotted ellipse
shows the single-detector mass metric obtained by using the harmonic sum
\ac{PSD}.}
\label{fig:massmetricvar}\end{figure}

Since marginalizing the coherent \ac{SNR} results in the creation of a single
virtual detector, this motivates the construction of an average mass metric
using this virtual detector's harmonic sum \ac{PSD}. In
Fig.~\ref{fig:massmetricvar}, we compare this virtual detector's mass metric to
a numerically-marginalized metric (i.e., the metric of
\eqref{eq:metricvolfullmarg}) projected down to the mass subspace. We find
excellent agreement between these two metrics.

It is interesting to note that the coherent \ac{SNR} is numerically equal to
the sum of squares of the \acp{SNR} that are obtained from the detectors
individually~\cite{Harry2011}, which is how the network \ac{SNR} is computed
for coincident searches. Because of this, we propose that when a single
template bank is desired for use in coincident searches, either the mass metric
(or inner products) used in the template placement algorithm be computed with
the harmonic sum \ac{PSD}.

It should be noted that when one is constructing a mass metric for a single set
of sky parameters, this introduces Dirac delta-functions, $\delta(\RA - \RA_0)$
and $\delta(\dec - \dec_0)$, into the marginalization integrals of the detector
responses. This leads us to use a weighted harmonic sum \ac{PSD} for the
virtual detector's \ac{PSD}, where the weighting for detector $Y$ is
$\left(F^Y_+(\RA_0, \dec_0)\right)^2 + \left(F^Y_\times(\RA_0, \dec_0)\right)^2$.

\subsection{Sky metric}
\label{sec:skymetric}

Just as we would like to create a sky-location-independent metric for the mass
subspace, we would like to create a mass-parameter-independent metric for the
sky-location subspace. From \cite{Fairhurst:2009tc, Wen:2010cr,
Fairhurst:2010is, Fairhurst:2012tf}, it has been found that the sky
localization accuracy associated with a network of \ac{GW} detectors is
dependent on the signal-weighted bandwidth of the detectors and the physical
separation of the detectors.

For short-duration, nonprecessing inspiral signals expanded to Newtonian order
in the amplitude, the power of the signal is given by $|h|^2 \propto f^{-7/3}$,
which is independent of the mass of the source objects. The relevant part of
the waveform that does depend on the mass of the source objects is the upper
frequency cutoff associated with the termination of the waveform, which is
inversely proportional to the total mass of the binary. This implies that the
signals with the lowest total masses will have the largest signal-weighted
bandwidth, and thus the best sky-localization accuracy. Because of this, we
propose to construct a single sky-space template bank for the particular mass
space given by the computing the marginalized metric from the lowest total mass
binary.

\begin{figure}
\includegraphics[]{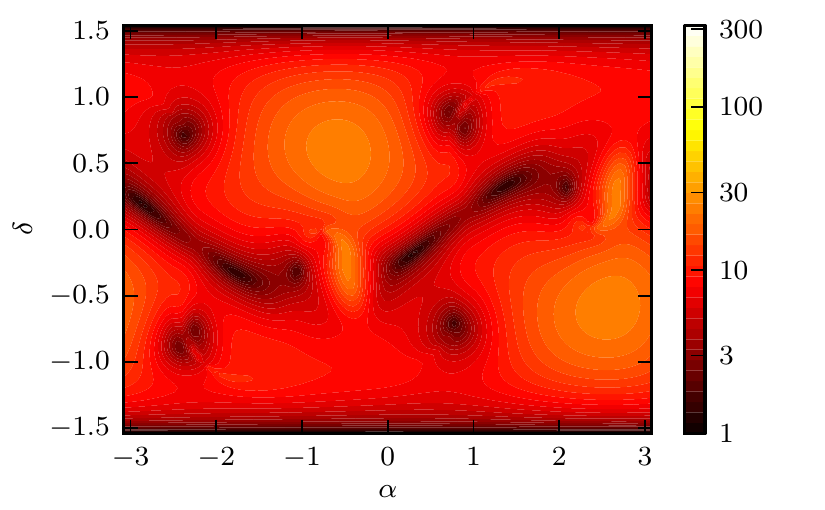}
\caption{We show the metric density as a function of the sky parameters for the
marginalized metric where the mass subspace has been projected out. This metric
was computed for the HLV network of advanced \ac{GW} detectors.}
\label{fig:margskymetric}\end{figure}

\begin{figure}
\includegraphics[]{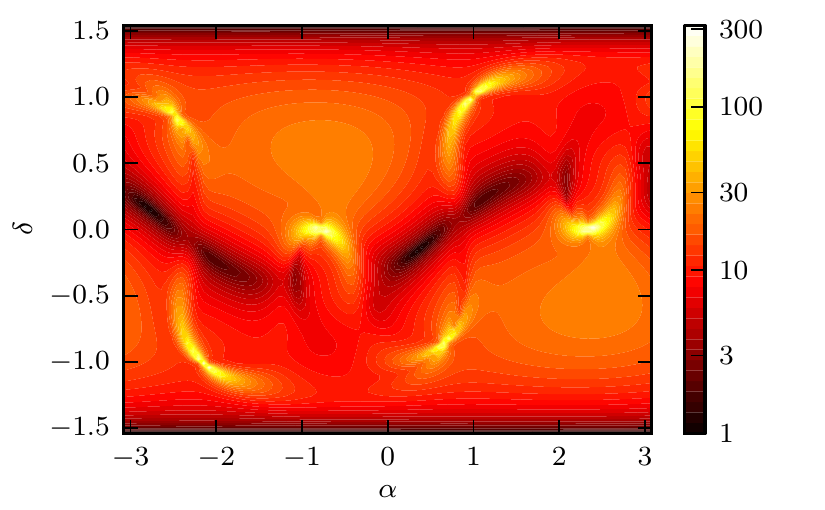}
\caption{We show the metric density as a function of the sky parameters for the
extrema-averaged metric where the mass subspace has been projected out. This
metric was computed for the HLV network of advanced \ac{GW} detectors.}
\label{fig:aveskymetric}\end{figure}

\begin{figure}
\includegraphics[]{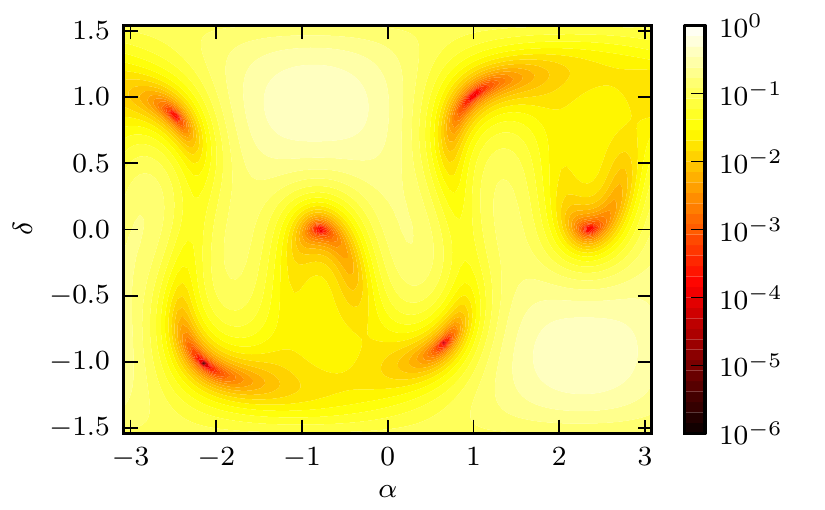}
\caption{We show $D$, \eqref{eq:D}, for the HLV network of \ac{GW}
detectors. The patterns of small $D$ seen here overlap with the patterns of
large metric density in Fig.~\ref{fig:aveskymetric}. This is due to the fact
that $D$ is in the denominator of the extrema-averaged metric.}
\label{fig:D}\end{figure}

In Fig.~\ref{fig:margskymetric} we show the variation of the marginalized sky
metric density associated with the HLV network of detectors as a function of
the source sky position, assuming the aLIGO~\cite{advLIGO} and
Adv.~Virgo~\cite{advVirgoPSD} \acp{PSD} and a signal from a \ac{BNS} system
with $m_1 = m_2 = 1.4 M_{\odot}$. This should be compared with
Fig.~\ref{fig:aveskymetric}, which shows the extrema-averaged sky metric
density. We see that there is more variation (and higher metric densities)
associated with this metric than with the marginalized metric. To understand
this, let us look at the form of \eqref{eq:avefstatmetric}, and in particular
concentrate on the term in the denominator, recalling that $D$ is given by
\eqref{eq:D}.  Figure~\ref{fig:D} shows how $D$ varies as a function of sky
location.  We see that points for which $D$ becomes near singular exhibit the
largest metric densities of Fig.~\ref{fig:aveskymetric}.

A qualitative understanding of why this occurs can be obtained by studying the
network response in a different coordinate system. In particular, if we
introduce an additional polarization angle between the geocentric frame and the
radiation frame, we find that choosing a particular value for this angle can
maximize the network response to one polarization, typically chosen to be the
new plus polarization. This is called the transformation to the \emph{dominant
polarization} frame~\cite{Klimenko:2005xv, Klimenko2006, Sutton:2009gi,
Harry2011, Bose:2011km}.  In particular, the detector polarization responses
for detector $Y$ are transformed as
\begin{gather}
F_{+}^{\mathrm{DP},Y}= F_{+}^Y \cos 2\chi^{\mathrm{DP}} + F_{\times}^Y \sin
2\chi^{\mathrm{DP}}, \\
F_{\times}^{\mathrm{DP},Y} = -F_{+}^Y \sin 2\chi^{\mathrm{DP}} + F_{\times}^Y
\cos 2\chi^{\mathrm{DP}},
\end{gather}
where 
\begin{equation}
\tan 4\chi^{\mathrm{DP}} = \frac{2 C}{A-B}.
\end{equation}
This choice of angle causes $C^{\mathrm{DP}}=0$, which diagonalizes the maximum
likelihood ratio matrix $\M$. The network response power of the two different
polarizations in this frame is found to be
\begin{gather}
A^{\mathrm{DP}} = \frac{1}{2}\left(A+B + \sqrt{\left(A+B\right)^2 - 4D}\right),
\\
B^{\mathrm{DP}} = \frac{1}{2}\left(A+B - \sqrt{\left(A+B\right)^2 - 4D}\right),
\end{gather}
which yields a total network response power of $A+B$. When $4 D \ll (A+B)^2$,
these approximate to
\begin{gather}
A^{\mathrm{DP}} \approx A + B - \frac{D}{A+B}, \\
B^{\mathrm{DP}} \approx \frac{D}{A+B},
\end{gather}
Thus we see that $D/(A+B)$ can be understood to be the network response power
of the subdominant polarization. This implies that the extrema-averaged metric
is dominated by locations where the network is sensitive to essentially only
one polarization.

\begin{figure*}[ht]
\subfigure[]{\includegraphics[]{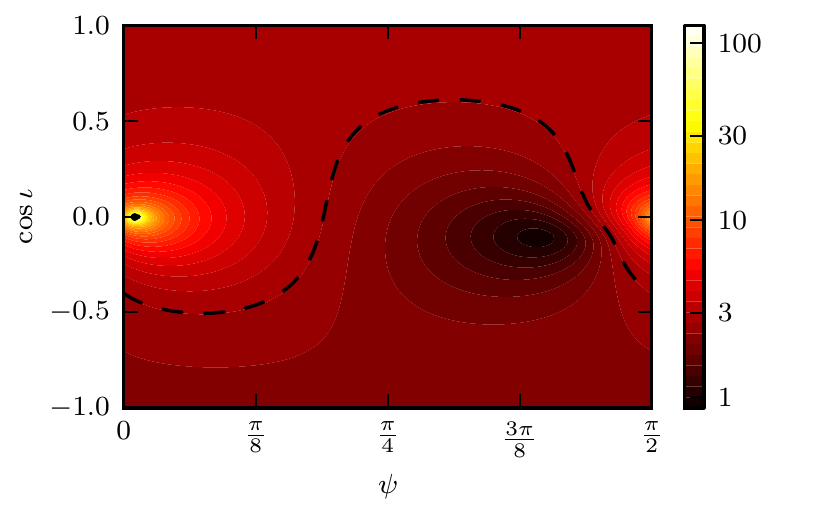}\label{subfig:rootdetg}}
\subfigure[]{\includegraphics[]{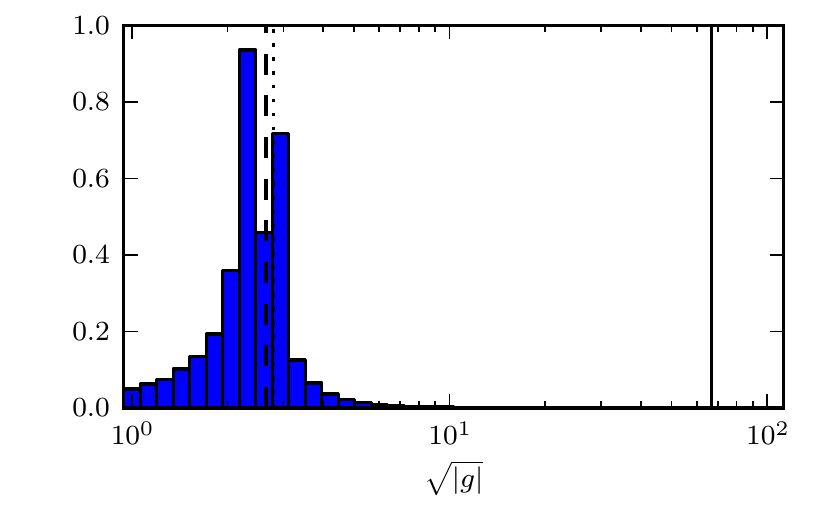}\label{subfig:rootdetghist}}
\subfigure[]{\includegraphics[]{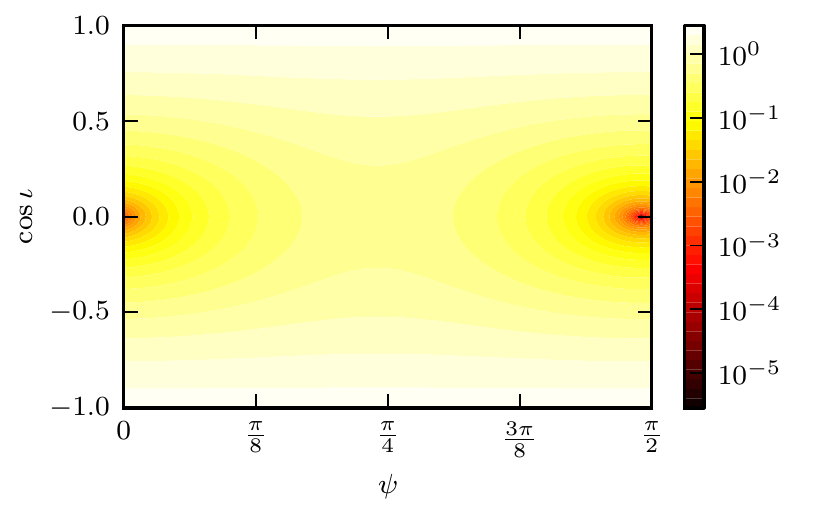}\label{subfig:domsnrs}}
\subfigure[]{\includegraphics[]{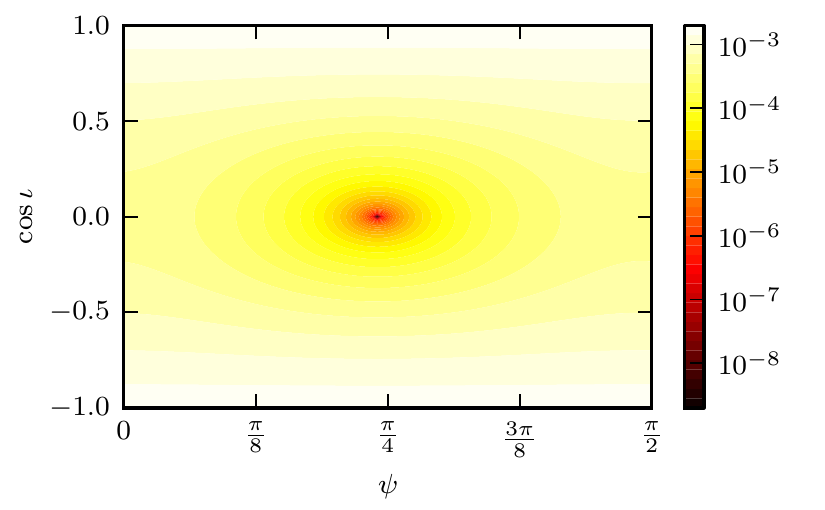}\label{subfig:subdomsnrs}}
\subfigure[]{\includegraphics[]{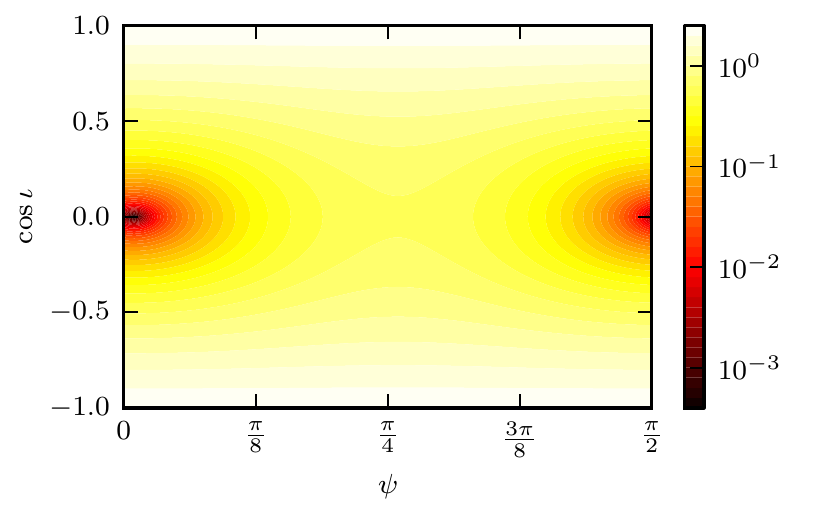}\label{subfig:snrs}}
\subfigure[]{\includegraphics[]{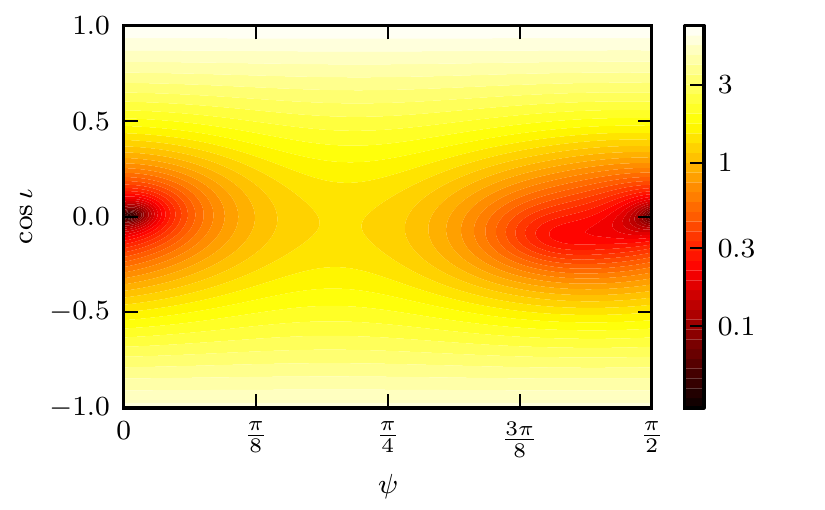}\label{subfig:gamma}}
\caption{\subref{subfig:rootdetg} shows the amplitude dependent metric density,
$\sqrt{|g|}$, associated with the sky parameters for a sky location where $D$
is near singular as a function of $\cos \iota$ and $\psi$. The solid (dashed)
contour shows the density of the extrema-averaged (marginalized) metric.
\subref{subfig:rootdetghist} shows a normalized histogram of $\sqrt{|g|}$. The
vertical solid and dashed lines again show the density of the extrema-averaged
metric and marginalized metric, respectively.  In addition, the vertical dotted
line shows the average value of $\sqrt{|g|}$.  \subref{subfig:domsnrs},
\subref{subfig:subdomsnrs}, and \subref{subfig:snrs} shows the coherent
\ac{SNR} associated with the dominant polarization, subdominant polarization,
and their sum, respectively, as a function of $\cos \iota$ and $\psi$. It is
observed that the regions where $\sqrt{|g|}$ is large, the coherent \ac{SNR}
associated with the dominant polarization is small. \subref{subfig:gamma} shows
that for the density of the $\rho^2$-weighted metric (i.e., the integrand of
the marginalized metric), the coherent \ac{SNR} weighting overwhelm the regions
of large $\sqrt{|g|}$.}
\label{fig:metricvspsicosi}\end{figure*}

This is further investigated in Fig.~\ref{fig:metricvspsicosi} where we look at
the amplitude parameter dependent metric density for a sky direction where $D$
is almost singular.  We find that the amplitude parameter dependent metric
density is largest for combinations of amplitude parameters that produce the
smallest coherent \acp{SNR} associated with the dominant polarization. As
expected, the coherent \ac{SNR} weighting associated with the marginalized
metric overwhelms these regions of large metric density.

\section{Conclusion}

We have derived a marginalized metric that can be used for template placement
associated with coherent searches for short-duration nonprecessing inspiral
\ac{GW} signals in data from a network of \ac{GW} detectors. The marginalized
metric emphasizes combinations of parameters that are more detectable than
previously proposed average metrics. This will help to focus computational
resources on regions of the extrinsic parameter space from which \acp{GW} are
more likely to be detected. In addition, we have shown how the marginalized
metric can be effectively separated into two two-dimensional metrics associated
with the sky and mass subspaces, respectively.

The marginalized metric is shown to take the same form when performing
searches for signals from randomly oriented systems (associated with ``all-sky"
searches) and signals from (anti-)aligned systems (associated with searches
targeting \acp{GW} from the progenitors of short \acp{GRB}).

The metric for the mass subspace is found to be well approximated by a
single-detector metric that is based on the harmonic sum of the \acp{PSD} of
detectors in the network. This metric is equally valid for coincidence-based
searches and coherent searches.

The metric for the sky subspace is found based on using the largest bandwidth
signal from the parameter space. The marginalized sky metric is compared to the
extrema-averaged sky metric. High density regions of the sky associated with
the extrema-averaged sky metric are found to be associated with locations where
the network response to the subdominant polarization is nearly singular and the
network response to the dominant polarization is small. This involved computing
the network response to the dominant and subdominant polarizations in terms of
$\F$-statistic quantities. Indeed, for an example sky location, combinations of
the inclination angle and polarization angle that contribute to the high
density of the extrema-averaged sky metric are shown to have a small
contribution to the marginalized sky metric.

These results will be useful in constructing template banks for either
coincident or coherent searches for short-duration, nonprecessing inspiral
signals in \ac{GW} data from the second generation \ac{GW} detectors.

\acknowledgments

The author would like to thank Lisa and Moriah Keppel for all of their love and
support during the completion of this work. The author would also like to thank
Jolien Creighton, Badri Krishnan, Andrew Lundgren, and Reinhard Prix for many
useful discussions that this work is based upon and Sukanta Bose and Archana
Pai for their comments on this manuscript. The author is supported from the Max
Planck Society. This document has LIGO document number LIGO-P1300084.

\appendix

\section{Amplitude Parameters}
\label{app:ampparam}

The detector independent amplitude parameters associated with our model for
short-duration, nonprecessing inspiral \ac{GW} signals are given as
\begin{equation} \begin{aligned} \label{eq:ampparams}
\A^1 &:= \;\;\; \mathscr{A}_+ \cos{\phi_0} \cos{2\psi} - \mathscr{A}_\cross
\sin{\phi_0} \sin{2\psi}, \\
\A^2 &:= \;\;\; \mathscr{A}_+ \cos{\phi_0} \sin{2\psi} + \mathscr{A}_\cross
\sin{\phi_0} \cos{2\psi}, \\
\A^3 &:= -\mathscr{A}_+ \sin{\phi_0} \cos{2\psi} - \mathscr{A}_\cross
\cos{\phi_0} \sin{2\psi}, \\
\A^4 &:= -\mathscr{A}_+ \sin{\phi_0} \sin{2\psi} + \mathscr{A}_\cross
\cos{\phi_0} \cos{2\psi},
\end{aligned} \end{equation}
where the polarization amplitudes $\mathscr{A}_+$ and $\mathscr{A}_\cross$ are
given by
\begin{equation} \begin{aligned} \label{eq:polamps}
\mathscr{A}_+ := \frac{h_0}{2}(1 + \cos^2{\iota}), \;
\mathscr{A}_\cross := h_0 \cos{\iota}.
\end{aligned} \end{equation}

\section{Basis Waveforms}
\label{app:basiswaveforms}

The four detector-dependent polarization-weighted basis waveforms
$\{h_{\mu}^{\Y}(t)\}$ are defined to be
\begin{equation} \begin{aligned}
h_1^\Y(t) &:= F_+^{\Y}(t - t^\Y) h_c(t - t^\Y), \\
h_2^\Y(t) &:= F_\cross^{\Y}(t - t^\Y) h_c(t - t^\Y), \\
h_3^\Y(t) &:= F_+^{\Y}(t - t^\Y) h_s(t - t^\Y), \\
h_4^\Y(t) &:= F_\cross^{\Y}(t - t^\Y) h_s(t - t^\Y),
\end{aligned} \end{equation}
where $F_{+}^{\Y}$ and $F_{\times}^{\Y}$ are the responses of detector $\Y$ to the
plus and cross polarization waveforms, respectively. For signals that satisfy
the long wavelength limit approximation~\cite{Rakhmanov2008}, these can be
defined as the double contraction of two tensors~\cite{PrixFstatTechNote},
\begin{equation}\label{eq:FpFx}
F_+^{\Y}(t) := \epsilon_+^{i j} d^\Y_{i j}(t), \;
F_\cross^{\Y}(t) := \epsilon_\cross^{i j} d^\Y_{i j}(t), \end{equation}
where $d^\Y_{i j}(t)$ is the \emph{detector response tensor} and
$\{\epsilon_{+,\cross}^{i j}\}$ are the \emph{polarization-independent basis
tensors of the radiation frame}. For an interferometric detector, the detector
response tensor is given by
\begin{equation}
d^\Y_{i j}(t) = \frac{1}{2} \left\{ \hat{l}^\Y_1(t) \otimes \hat{l}^\Y_1(t) -
\hat{l}^\Y_2(t) \otimes \hat{l}^\Y_2(t) \right\}_{i j},
\end{equation}
where $\hat{l}^\Y_1$ $(\hat{l}^\Y_2)$ is the unit vector pointing along
interferometer~\Y's first (second) arm away from the interferometer's vertex.
As defined in~\cite{Keppel:2012ye, PrixFstatTechNote}, the
polarization-independent basis tensors can be built from the basis vectors of
the radiation frame $\{\hat{\xi}, \hat{\eta}, -\hat{n}\}$, where $-\hat{n}$ is
the direction of propagation, and $\{\hat{\xi}, \hat{\eta}\}$ are basis vectors
in the wave-plane (i.e., the plane perpendicular to direction of propagation).
The basis vectors $\hat{\xi}$ and $\hat{\eta}$ can be defined with respect to
$\hat{n}$ as
\begin{align}
\hat{\xi} := \frac{\hat{n} \cross \hat{z}}{\abs{\hat{n} \cross \hat{z}}}, \;
\hat{\eta} := \hat{\xi} \cross \hat{n}.
\end{align}
In a fixed reference frame centered at the geocenter, where
\begin{equation}
\hat{n} = (\cos{\dec} \cos{\RA}, \cos{\dec} \sin{\RA}, \sin{\dec}),
\end{equation}
the wave-plane basis vectors are
\begin{gather}
\hat{\xi} = (\sin{\RA}, -\cos{\RA}, 0), \\
\hat{\eta} = (-\sin{\dec} \cos{\RA}, -\sin{\dec} \sin{\RA}, \cos{\dec}).
\end{gather}
Using these conventions, the polarization-independent basis tensors are given
as
\begin{equation} \begin{aligned}
\epsilon^{i j}_+ &:= \left\{\hat{\xi} \otimes \hat{\xi} - \hat{\eta} \otimes
\hat{\eta}\right\}^{i j}, \\
\epsilon^{i j}_\cross &:= \left\{\hat{\xi} \otimes \hat{\eta} + \hat{\eta}
\otimes \hat{\xi}\right\}^{i j}.
\end{aligned} \end{equation}
%

\section{Amplitude Parameter Integrals}
\label{app:ampaves}

Here we compute integrals of products of the amplitude parameters over the
physical parameters that compose them. Since these amplitude parameters are
used in conjunction with either $\M_{\mu \nu}$ or $\G_{\mu \nu i j}$ summed
over the free indices, and since $\M_{\mu \nu}$ and $\G_{\mu \nu i j}$ have the
same symmetries, we make use of those symmetries to compute only the necessary
combinations of products of the amplitude parameters. Those combinations are
\begin{align}\label{eq:ampparamcombos}
(\A^1 \A^1 + \A^3 \A^3) &= \mathscr{A}_+^2 \cos^2(2\psi) + \mathscr{A}_\times^2
\sin^2(2\psi), \\
(\A^2 \A^2 + \A^4 \A^4) &= \mathscr{A}_+^2 \sin^2(2\psi) + \mathscr{A}_\times^2
\cos^2(2\psi), \\
(\A^1 \A^2 + \A^3 \A^4) &= (\mathscr{A}_+^2 - \mathscr{A}_\times^2)
\cos(2\psi)\sin(2\psi), \\
(\A^1 \A^4 - \A^2 \A^3) &= \mathscr{A}_+ \mathscr{A}_\times.
\end{align}
We are interested in integrals of these quantities of the form
\begin{equation}
\int_0^{2\pi} \int_0^{2\pi} \int_0^{\pi} (\cdot) \sin\iota \diff\phi_0
\diff\psi \diff\iota,
\end{equation}
where $(\cdot)$ refer to one of the combinations of amplitude parameters in
\eqref{eq:ampparamcombos}. Performing these integrals for each of the
combinations of amplitude parameters above, we find that the only nonzero
combinations are
\begin{multline}\label{eq:ampfullint}
\int_0^{2\pi} \int_0^{2\pi} \int_0^{\pi} (\A^1 \A^1 + \A^3 \A^3) \sin\iota
\diff\phi_0 \diff\psi \diff\iota = \\
\int_0^{2\pi} \int_0^{2\pi} \int_0^{\pi} (\A^2 \A^2 + \A^4 \A^4) \sin\iota
\diff\phi_0 \diff\psi \diff\iota = \frac{8 \pi^2 h_0^2}{5}.
\end{multline}

If we assume the orbital angular momentum vector is parallel to the line of
site, as is the case for when we are searching for signals associated with
\acp{GRB}, we obtain
\begin{align}\label{eq:ampintpsi1}
\int_0^{2\pi} \int_0^{2\pi} (\A^1 \A^1 + \A^3 \A^3) |_{\iota = \iota_0}
\diff\phi_0 \diff\psi &= \nonumber \\
\int_0^{2\pi} \int_0^{2\pi} (\A^2 \A^2 + \A^4 \A^4) |_{\iota = \iota_0}
\diff\phi_0 \diff\psi &= 4\pi^2 h_0^2, \\
\int_0^{2\pi} \int_0^{2\pi} (\A^1 \A^4 - \A^2 \A^3) |_{\iota = \iota_0}
\diff\phi_0 \diff\psi &= \pm 4\pi^2 h_0^2,
\end{align}
where the $\pm$ in the third term is associated with inclination angles of
$\iota_0 = \{0, \pi\}$.  We assume is it equally likely that when we observe a
\ac{GRB}, the orbital angular momentum is pointing towards us or away from us,
thus the last term above will average to zero.  Since the same average
combinations of the amplitude parameters are non-zero as in the unknown
inclination angle case and the same average values for the combinations of
marginalized parameters will be found in the numerator and denominator of the
marginalized metric, the metric used for searching for \ac{GW} signals associated
with \acp{GRB} is the same as in the unknown inclination angle case.

\bibliography{references}
\end{document}